\documentstyle[sprocl,epsfig,amsmath]{article}

\def\NIMA{{\em Nucl. Instrum. Methods} A}

\def\be{\begin{equation}}
\def\ee{\end{equation}}
\def\bea{\begin{eqnarray}}
\def\eea{\end{eqnarray}}
\begin{document}

\thispagestyle{empty}

\begin{flushright}
{\large \tt LC-DET-2004-015}\\[20mm]
\end{flushright}

\begin{center}
    {\LARGE \bf High Magnetic Field Performance\\ of a GEM-TPC}\\[10mm]
    {\normalsize \sc Peter Wienemann}\\[5mm]
    {\normalsize \it Deutsches Elektronen-Synchrotron DESY,\\[0.2ex]
     Notkestr.~85, 22607 Hamburg, Germany}
\end{center}

\vspace{30mm}

\begin{abstract}
  \noindent
  A large volume time projection chamber (TPC) has been proposed as main
  tracking device for a detector at a future linear $\text{e}^+ \text{e}^-$
  collider. Gas electron multipliers (GEMs) are studied as potential
  replacement of the conventional wire based gas amplification system of TPCs.
  This talk presents recent results from R\&D activities with a small GEM-TPC
  prototype. The spatial resolution was measured for different magnetic
  fields up to 4 T.
\end{abstract}

\vspace{20mm}

\begin{center}
   {\normalsize Proceedings of the International Conference on Linear Colliders (LCWS 04)\\
                Paris, April 19-23, 2004 }
\end{center}

\newpage

\title{HIGH MAGNETIC FIELD PERFORMANCE OF A GEM-TPC\\}

\setcounter{page}{1}

\author{PETER WIENEMANN}

\address{Deutsches Elektronen-Synchrotron DESY,\\
         Notkestr.~85, 22607 Hamburg, Germany}

%%%%%%%%%%%%%%%%%%%%%%%%%%%%%%%%%%%%%%%%%%%%%%%%%%%%%%%%%%%%%%
% You may repeat \author \address as often as necessary      %
%%%%%%%%%%%%%%%%%%%%%%%%%%%%%%%%%%%%%%%%%%%%%%%%%%%%%%%%%%%%%%

\maketitle\abstracts{
  A large volume time projection chamber (TPC) has been proposed as main
  tracking device for a detector at a future linear $\text{e}^+ \text{e}^-$
  collider. Gas electron multipliers (GEMs) are studied as potential
  replacement of the conventional wire based gas amplification system of TPCs.
  This talk presents recent results from R\&D activities with a small GEM-TPC
  prototype. The spatial resolution was measured for different magnetic
  fields up to 4 T.
}

%***********************************************************************
\section{Introduction}
%***********************************************************************
\label{sec:introduction}

The ambitious physics program of a future linear $\text{e}^+
\text{e}^-$ collider poses stringent requirements on the precision of
its tracker as part of a precise overall detector. The measurement of
the Higgs properties e.~g.~requires excellent momentum resolution for
mass reconstruction and good $dE/dx$ accuracy for particle
identification for branching ratio measurements. A large volume time
projection chamber (TPC) is considered a good candidate as a tracking
device for this detector. Contrary to conventional TPCs with a
multiwire proportional chamber (MWPC) technique for gas amplification,
future TPCs are likely to make use of micro pattern gas detectors
(MPGDs).  The best known representatives of such MPGDs are the gas
electron multiplier (GEM) \cite{ref:GEM} and micromegas
\cite{ref:micromegas}.  MPGDs have amplification structures of order
100 $\mu$m giving rise to only tiny $\vec{E} \times \vec{B}$ effects,
they provide a fast and narrow electron signal and have intrinsic ion
feedback suppression -- all features making them good candidates for a
gas amplification system in TPCs.

%***********************************************************************
\section{The DESY TPC Prototype}
%***********************************************************************

To study the potential of TPCs with GEM foils for gas amplification, a small
TPC prototype has been built at DESY. The chamber has a length of 800~mm with
a diameter of 270~mm. The diameter has been chosen such that the chamber fits
into a superconducting 5 T magnet with a 280~mm aperture available for
detector R\&D at DESY. The chamber is equipped with 24 $\times$ 8 $=$ 192
readout pads of size 2.2 $\times$ 6.2~mm$^2$. The maximal drift distance is
680~mm. Gas amplification is provided by a triple GEM structure with two 2~mm
transfer gaps and a 3~mm induction gap. The readout of the chamber is based
on electronics built for the ALEPH experiment at LEP.

%***********************************************************************
\section{Single Point Resolution in High Magnetic Fields}
%***********************************************************************

One of the important questions which are addressed by the present TPC R\&D
activities is the achievable single point resolution of a TPC with a GEM based
amplification system in magnetic fields as high as 4 T as proposed in the
technical design report for TESLA \cite{ref:TeslaTDR}. The single point
resolution is a useful number to compare the resolutions obtained from
different prototypes and to extrapolate from small prototypes to the
performance of a large scale device.

To find an answer to this question a number of measurements with
cosmic muons was carried out in the 5 T magnet at DESY. Data were
taken at magnetic fields of 0~T, 0.75~T, 1~T and 4~T. The
reconstruction of the track parameters from the data is done in three
steps. First, in each of the planes spanned by a pad row and the drift
direction, clusters are reconstructed from the hits of the various
channels. The $x$ coordinate, describing the position along a pad row,
is obtained using a center-of-gravity method. For each pad row the
vertical position of the pad center is assigned to the $y$ coordinate.
The $z$ coordinate of a cluster, describing the drift distance, is set
to the position of the inflection point of the rising edge of the
corresponding cluster.  In a second step the reconstructed
three-dimensional points are combined into tracks using a
three-dimensional track following algorithm. Finally track parameters
are fitted to the points belonging to the various tracks using a
$\chi^2$ technique.  The track parameters are fitted independently in
the $xy$ and the $yz$ plane. The fitted parameters are the
intersection of the track with the $xz$ plane at $y = 0$, the slope in
the $xy$ and $yz$ plane and the curvature in the $xy$ plane.

To determine the single point resolution, the following procedure is
used: Remove the track point in row $i$ and re-fit the $x$ axis
intercept of the track without point $i$. The distance of point $i$ to
the re-fitted track is called residual for row $i$.  The resolution is
obtained from the standard deviation of the residual distribution of
the four central rows for the whole track sample.

\begin{figure}[ht]
\begin{center}
    \epsfig{file=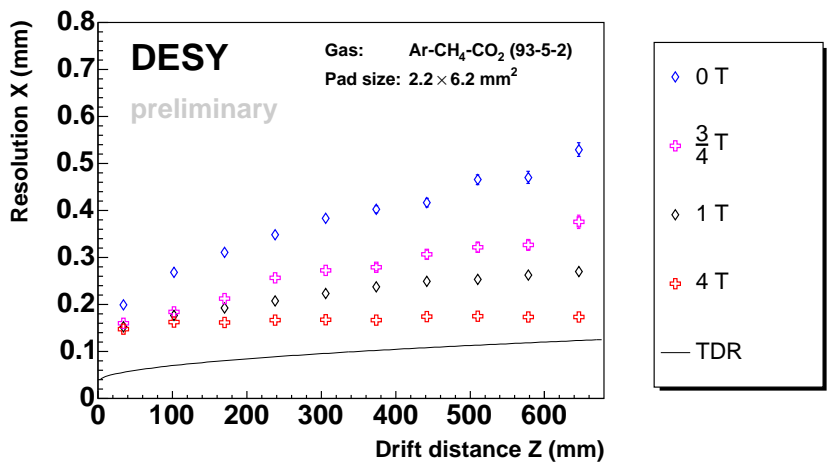,width=0.99\textwidth}
\end{center}
\caption{The single point resolution versus drift distance for different
  magnetic fields $^4$.}
\label{fig:resolution}
\end{figure}

Fig.~\ref{fig:resolution} shows the transverse single point resolution
versus drift distance obtained with this technique for the gas mixture
Ar-CH$_4$-CO$_2$ (93-5-2). For the data collected up to 1 T, an
increase of the resolution with drift distance is observed. In these
cases the spatial resolution is limited by the diffusion in the drift
region of the TPC. At 4 T the situation is different. Up to the
maximal drift distance of 680 mm the resolution is flat as a function
of the drift distance. With the present preliminary analysis a
resolution of $\approx$ 160 $\mu$m is achieved which is not yet
sufficient to fulfill the TESLA technical design report
specifications. This indicates that the magnetic field suppresses the
diffusion in the drift region to an extent that it is not the limiting
factor anymore.  Possible reasons might be either an inappropriate
ratio of the charge distribution width to the pad width or analysis
systematics. Further investigations are ongoing to understand the
measurements.

%***********************************************************************
\section{Conclusion}
%***********************************************************************

A small TPC prototype with GEM foils for gas amplification has been
successfully built to measure the single point resolution in high
magnetic fields. Cosmic muon runs were carried out in B fields up to 4
T. A transverse resolution of 160 $\mu$m has been achieved with the
present preliminary analysis.  Further studies are under way to
understand the limiting factors and to improve the performance.

%***********************************************************************
\section*{References}
%***********************************************************************


\begin{thebibliography}{99}

\bibitem{ref:GEM} F.~Sauli, \NIMA, \rm {\bf 386} (1997) 531.
  
\bibitem{ref:micromegas} Y.~Giomataris {\it et.~al.}, \NIMA, \rm {\bf 376}
  (1996) 29.
  
\bibitem{ref:TeslaTDR} T.~Behnke, S.~Bertolucci, R.-D.~Heuer, and
  R.~Settles, TESLA Technical Design Report, DESY, Hamburg, Germany,
  DESY 2001-011 and ECFA 2001-209 (2001).

\bibitem{ref:MatthiasThesis} M.~E.~Janssen, Aufl\"osungsstudien an einer
  Zeit-Projektions-Kammer mit GEM-Auslese, Diploma thesis, University of
  Dortmund (2004).

\end{thebibliography}
\end{document}